\documentclass[english,aps, superscriptaddress, floatfix, 12pt]{revtex4}
\usepackage[T1]{fontenc}
\usepackage[latin1]{inputenc}
\usepackage{graphicx}
\usepackage{amssymb}
\usepackage{epsfig}
\usepackage{amsmath}
\usepackage{psfrag}

\usepackage{babel}
\makeatother
\begin{document}

\title{Shear viscosity of a nonperturbative gluon plasma}

\author{Dmitri Antonov\\
{\it Departamento de F\'isica and Centro de F\'isica das Interac\c{c}\~oes Fundamentais,}\\ 
{\it Instituto Superior T\'ecnico, UT Lisboa,
Av. Rovisco Pais, 1049-001 Lisboa, Portugal}}

\noaffiliation

\begin{abstract}
Shear viscosity is evaluated within a model of the gluon plasma, which is based entirely on the stochastic nonperturbative fields. We consider two types of excitations of such fields,
which are characterized by the thermal correlation lengths $\sim (g^2T)^{-1}$ and $\sim (g^4T)^{-1}$, where $g$ is the finite-temperature Yang--Mills coupling. Excitations of the first type correspond to the genuine 
nonperturbative stochastic Yang--Mills fields, while excitations of the second type 
mimic the known result for the shear viscosity of the perturbative Yang--Mills plasma.
We show that the excitations of the first type produce only an 
${\cal O}(g^{10})$-correction to this result.
Furthermore, a possible interference between excitations of these two types yields a somewhat larger, 
${\cal O}(g^7)$, correction to the leading perturbative Yang--Mills result.
 
Our analysis is based on the Fourier transformed Euclidean Kubo formula, which represents an integral equation for the shear spectral density.
This equation is solved by seeking the spectral density in the form of the 
Lorentzian Ans\"atze, whose widths are defined by the two thermal 
correlation lengths and by their mean value, which corresponds to the said interference between the two types of excitations. Thus, within one and the same formalism,
we reproduce the known result for the shear viscosity 
of the perturbative Yang--Mills plasma, and account for possible nonperturbative corrections to it.
\end{abstract}

\maketitle

\section{Introduction} 
Over the last ten years, it has been widely discussed that the quark-gluon plasma produced 
in the RHIC experiments can resemble an almost perfect quantum liquid, which is characterized
by the values of the shear-viscosity to the entropy-density ratio, $\eta/s$, much smaller than unity~\cite{es}.
Comparison with the empirical data for
water, helium, and nitrogen shows that their $(\eta/s)$-ratios reach minima in the vicinity of the corresponding 
liquid-gas phase transitions~\cite{ckm}. Given different types of phase 
transitions and different types of molecules for the above-mentioned three substances, one can guess 
that such a temperature-behavior of $\eta/s$ is quite general. Using this observation,
one can naturally assume that for the quark-gluon plasma
the minimum of the $(\eta/s)$-ratio is reached in the vicinity of the 
deconfinement phase transition. This minimum can be set to the value of $1/(4\pi)$, obtained within 
$({\cal N}=4)$ supersymmetric Yang--Mills theory, which was suggested as the 
lower bound for the $(\eta/s)$-ratio~\cite{pss}. With the increase of temperature, $\eta/s$ is expected to rise from this bound up to the values predicted by the perturbative-QCD calculations~\cite{amy}. 
Thus, a problem can be posed as how to model these essential features in the temperature-behavior 
of the $(\eta/s)$-ratio.

In the present Letter, we address this issue for the purely gluonic plasma, within a model based entirely on the stochastic nonperturbative fields. In particular, we manage to reproduce the said high-temperature behavior of $\eta/s$ in perturbative Yang--Mills theory by imposing the correlation length of these fields to be $\sim (g^4T)^{-1}$, where $g$ is the finite-temperature Yang--Mills coupling. This correlation length is recognizable 
as the mean time needed for a parton undergoing Coulomb scatterings in the 
gluon plasma to deflect by an angle of the order of unity~\cite{arn}. Of course, 
besides the ultrasoft momentum scale $\sim g^4T$, stochastic nonperturbative fields possess just the soft scale $\sim g^2T$, which defines the high-temperature behavior of such quantities as the spatial string 
tension~\cite{f1} and the nonperturbative gluonic condensate~\cite{nik}. 
Hence, 
the key ingredient of our model is the presence, in the 
deconfinement phase $(T>T_c)$ of interest, of the two types of excitations of the nonperturbative 
fields. These excitations are characterized by the parametrically different correlation lengths,
$\sim (g^2T)^{-1}$ and $\sim (g^4T)^{-1}$.
Excitations of the first type describe an extrapolation of the genuine nonperturbative stochastic vacuum Yang--Mills fields to the deconfinement phase~\cite{nik}. Rather, excitations of the second type 
are introduced with the purpose to mimic the known perturbative 
contribution~\cite{amy} to the shear viscosity. 

The goal of the present study is therefore twofold:
to quantify the relative contribution to $\eta/s$, which is produced by the genuine nonperturbative fields [i.e. those with the correlation length $\sim (g^2T)^{-1}$] with respect to the known perturbative 
contribution, and to evaluate the contribution to $\eta/s$ produced by the perturbative-nonperturbative interference. These two issues will be addressed by obtaining 
the shear spectral density from an integral equation given by the Fourier transformed Kubo formula. That will be done by seeking the spectral density 
as a superposition of the Lorentzian Ans\"atze. As a result, we find that the widths of these Lorentzians are given by the said momentum scales, $\sim g^2T$ and $\sim g^4T$, as well as by their mean value (for the
interference between the perturbative and nonperturbative interactions). 

In the next Section, we perform the corresponding analytic and numerical calculations. In Section~III, we 
give a brief summary of the results obtained.

\section{Calculation of the $(\eta/s)$-ratio}

\noindent
The spectral density $\rho\equiv\rho(\omega,T)$, defining the shear viscosity $\eta\equiv\eta(T)$ as 
\begin{equation}
\label{0}
\left.\eta=\pi\frac{d\rho}{d\omega}\right|_{\omega=0},
\end{equation}
can be obtained from the following Euclidean Kubo formula~\cite{kw}:
\begin{equation}
\label{kubo} 
\int_0^\infty d\omega~ \rho~ \frac{\cosh\left[\omega\left(x_4-\frac{\beta}{2}
\right)\right]}{\sinh(\omega\beta/2)}=\int d^3x\sum\limits_{n=-\infty}^{+\infty}
U_T({\bf x},x_4+\beta n).
\end{equation}
Here, $\beta\equiv 1/T$, $n$ labels the winding mode, 
and $U_T$ is the finite-temperature correlation function of the $(1,2)$-component 
of the Yang--Mills energy-momentum tensor $\Theta_{\mu\nu}$:
\begin{equation}
\label{u}
U_T({\bf x},x_4+\beta n)\equiv \bigl<\Theta_{12}({\bf 0},0)\Theta_{12}({\bf x},x_4+\beta n)\bigr>_T,~~~ {\rm where}~~~
\Theta_{12}=g^2F_{1\mu}^aF_{2\mu}^a.
\end{equation}
The Kubo formula represents an integral equation for $\rho$. To solve this equation, we 
first Fourier transform it. This method of solving the equation is inspired by the observation that 
\begin{equation}
\label{2}
\frac{\cosh\left[\omega\left(x_4-\frac{\beta}{2}\right)\right]}{\sinh(\omega\beta/2)}=
2T\cdot \omega\sum\limits_{k=-\infty}^{+\infty}\frac{{\rm e}^{i\omega_kx_4}}{\omega^2+\omega_k^2},
\end{equation}
where $\omega_k=2\pi Tk$ is the $k$-th Matsubara frequency. One further notices that,
for nonperturbative fields at issue, $U_T$ exponentially falls off at a distance defined 
by the thermal correlation 
length of those fields. For this reason, we consider the maximally general exponential Ansatz for $U_T$, 
which is provided by the MacDonald functions. Namely, we start with the 
following sum, which generalizes the one on the right-hand side of Eq.~(\ref{2}):
$$S\equiv\sum\limits_{k=-\infty}^{+\infty}\frac{{\rm e}^{i\omega_k x_4}}{(\omega_k^2+m^2)^\alpha}=
\frac{1}{\Gamma(\alpha)}\int_0^\infty d\lambda{\,}\lambda^{\alpha-1}{\rm e}^{-\lambda m^2}
\sum\limits_{k=-\infty}^{+\infty}{\rm e}^{-\lambda\omega_k^2+i\omega_k x_4}.$$
Here, $\Gamma(\alpha)$ stands for the Gamma-function,
$\alpha>0$, and $m=m(T)$ is some mass parameter. The sum over Matsubara modes $k$ can be transformed into a 
sum over winding modes $n$, yielding the following intermediate result:
$$S=\frac{\beta}{2\sqrt{\pi}\Gamma(\alpha)}\int_0^\infty d\lambda{\,}\lambda^{\alpha-\frac32}
{\rm e}^{-\lambda m^2}\sum\limits_{n=-\infty}^{+\infty}{\rm e}^{-\frac{(x_4+\beta n)^2}{4\lambda}}.$$
One can further multiply this expression by $1=(4\pi\lambda)^{-3/2}\int d^3x{\,}{\rm e}^{-
\frac{{\bf x}^2}{4\lambda}}$, which yields
$$S=\frac{\beta}{16\pi^2\Gamma(\alpha)}\int_0^\infty d\lambda{\,}\lambda^{\alpha-3}
{\rm e}^{-\lambda m^2}\int d^3x\sum\limits_{n=-\infty}^{+\infty}{\rm e}^{-\frac{{\bf x}^2+
(x_4+\beta n)^2}{4\lambda}}.$$
Performing then the $\lambda$-integration, one obtains
$$S=\frac{\beta\cdot m^{4-2\alpha}}{2^{\alpha+1}\pi^2\Gamma(\alpha)}\int d^3x\sum\limits_{n=-\infty}^{+\infty}
\frac{K_{2-\alpha}(m\sqrt{{\bf x}^2+(x_4+\beta n)^2})}{(m\sqrt{{\bf x}^2+(x_4+\beta n)^2})^{2-\alpha}},$$
where $K_\nu(x)$ stands for the MacDonald function. 

Hence, we assume the correlation function~(\ref{u}) of the following form:
\begin{equation}
\label{uT}
U_T({\bf x},x_4+\beta n)=N(\alpha){\,}G_T^2{\,}
\frac{K_{2-\alpha}(m\sqrt{{\bf x}^2+(x_4+\beta n)^2})}{(m\sqrt{{\bf x}^2+(x_4+\beta n)^2})^{2-\alpha}},
\end{equation}
where $N(\alpha)$ is a numerical parameter, and $G_T\equiv\langle(gF_{\mu\nu}^a)^2\rangle_T$ is the 
finite-temperature nonperturbative gluonic condensate. Then the Fourier-transformed Kubo formula reads
\begin{equation}
\label{main}
\int_0^\infty d\omega{\,}\frac{\omega\rho}{\omega^2+\omega_k^2}=\pi^2 2^\alpha N(\alpha){\,}\Gamma(\alpha)
G_T^2{\,}\frac{m^{2\alpha-4}}{(\omega_k^2+m^2)^\alpha}.
\end{equation}
We use now for $\rho$ a Lorentzian Ansatz with the width $m$:
\begin{equation}
\label{ans}
\rho=\frac{C{\,}\omega}{\omega^2+m^2},
\end{equation} 
where $C=C(T)$ is the sought function of dimensionality (mass)$^5$. Notice that, although the 
asymptotic freedom requires $\rho\propto\omega^4$ at $\omega\gg T$ (cf. e.g. Ref.~\cite{me}),
it is the Lorentzian part of $\rho$ which matters for $\eta$, since it defines the derivative of 
$\rho$ at $\omega=0$. 
As it then follows from Eq.~(\ref{0}), the shear viscosity is expressed in terms of $C$ as 
$\eta=\pi C/m^2$. Substituting Ansatz~(\ref{ans}) into Eq.~(\ref{main}), we are left with the integral
$$\int_0^\infty d\omega{\,}\frac{\omega^2}{(\omega^2+\omega_k^2)(\omega^2+m^2)}=\frac{\pi}{2(|\omega_k|+m)}.$$
Setting $\alpha=1/2$, we arrive at the relation
$$C=\sqrt{8\pi^3}{\,}N{\,}\frac{G_T^2}{m^3}{\,}\frac{|\omega_k|+m}{\sqrt{\omega_k^2+m^2}},$$
where from now on $N=N(1/2)$. Thus, we see that, for $\alpha=1/2$, an exponentially 
falling off function~(\ref{uT}) is compatible with the Lorentzian Ansatz~(\ref{ans}) for $k=0$
(that is, in the high-temperature dimensionally-reduced theory) and for $|k|\gg 1$. With a given form~(\ref{2}) of the kernel in the integral equation~(\ref{kubo}), which is prescribed by the fluctuation-dissipation theorem, 
and with the use of the Lorentzian Ansatz,
a better accuracy can hardly be achieved. Thus, we use the formula
\begin{equation}
\label{eta}
\eta\simeq\sqrt{8\pi^5}{\,}N{\,}\frac{G_T^2}{m^5},
\end{equation}
which is supported by the observation that the correcting factor 
\begin{equation}
\label{f3}
\frac{|\omega_k|+m}{\sqrt{\omega_k^2+m^2}}
\end{equation} 
is equal to 1 for $T>T_{*}$, where $T_{*}$ is the temperature of dimensional reduction. By the end of 
our analysis, we will numerically evaluate maximum possible deviations of the correcting factor from 1,
which can take place for the temperatures $T\in(T_c,T_{*})$ at $|k|\sim 1$.

We proceed now to the calculation of the coefficient $N$.
To this end, we first notice that the $(T=0)$-counterpart of the function~(\ref{uT}) at $\alpha=1/2$ reads 
\begin{equation}
\label{uu}
U_0(x)=N{\,} G^2{\,} \frac{K_{3/2}(m_0|x|)}{(m_0|x|)^{3/2}},~~ 
{\rm where}~~ \frac{K_{3/2}(z)}{z^{3/2}}
=\sqrt{\frac{\pi}{2}}{\,}\frac{1}{z^2}{\,}\left(\frac1z+1\right){\,}{\rm e}^{-z},
\end{equation}
$G\equiv \langle(gF_{\mu\nu}^a)^2\rangle$, and the subscript ``0'' means ``at zero temperature''. Second, we
express this correlation function in terms of the 2-point functions 
of $F_{\mu\nu}^a$'s by using the so-called Gaussian-dominance hypothesis. This hypothesis, supported by the 
lattice simulations~\cite{lat}, 
states that the connected 4-point function of $F_{\mu\nu}^a$'s can be 
neglected compared to the pairwise products of the 2-point functions. For the 
function $U_0(x)=\langle g^4F_{1\mu}^a(0)F_{2\mu}^a(0)
F_{1\nu}^b(x)F_{2\nu}^b(x)\rangle$ at issue (cf. Eq.~(\ref{u})), the Gaussian-dominance hypothesis yields 
\begin{equation}
\label{u0}
U_0(x)\simeq\langle g^2F_{1\mu}^a(0)F_{1\nu}^b(x)\rangle \langle g^2F_{2\mu}^a(0)F_{2\nu}^b(x)\rangle
+\langle g^2F_{1\mu}^a(0)F_{2\nu}^b(x)\rangle \langle g^2F_{2\mu}^a(0)F_{1\nu}^b(x)\rangle,
\end{equation}
where we have taken into account that $\langle g^2F_{1\mu}^a(0)F_{2\mu}^a(0)\rangle=0$.
The contribution of stochastic nonperturbative fields to the function 
$\langle g^2F_{\mu\nu}^a(0)F_{\lambda\rho}^b(x)\rangle$ can with a high accuracy be parametrized 
as~\cite{lat, svm}
\begin{equation}
\label{p}
\langle g^2F_{\mu\nu}^a(0)F_{\lambda\rho}^b(x)\rangle=
\frac{G}{12}\left(\delta_{\mu\lambda}\delta_{\nu\rho}-\delta_{\mu\rho}\delta_{\nu\lambda}\right)\cdot
\frac{\delta^{ab}}{N_c^2-1}\cdot D(x).
\end{equation}
In Eq.~(\ref{p}),
the dimensionless function $D(x)$ exponentially falls off at a distance called the vacuum correlation length. 
Substituting Eq.~(\ref{p}) into Eq.~(\ref{u0}), one readily obtains
$U_0\simeq\frac{G^2}{72(N_c^2-1)}D^2$. Comparing this expression with Eq.~(\ref{uu}), and setting 
from now on $N_c=3$, we have 
\begin{equation}
\label{d}
D(x)=24{\,}\left[N{\,}\frac{K_{3/2}(m_0|x|)}{(m_0|x|)^{3/2}}\right]^{1/2}.
\end{equation}

As mentioned above, we consider two types of excitations of the nonperturbative 
fields, which are characterized by the correlation lengths
$\sim (g^2T)^{-1}$ and $\sim (g^4T)^{-1}$.
We start with the excitations of the first type. These excitations exist foremost in  
the confinement phase $(T<T_c)$, where they yield the string tension~\cite{svm} $\sigma=\frac{G}{144}\int d^2x D({\bf x})$ corresponding to the static sources in the fundamental representation.
On the other hand, the function $D(x)$ in the confinement phase is conventionally parametrized by just an 
exponent~\cite{lat, svm} which, once compared to Eq.~(\ref{d}), would be $D(x)={\rm e}^{-m_0|x|/2}$.
Plugging both this exponential parametrization and parametrization~(\ref{d}) into the said formula for 
$\sigma$, we obtain the corresponding coefficient $N$:
\begin{equation}
\label{N}
N=\frac{1}{\left[6\int_0^\infty dx{\,} x^{1/4}\sqrt{K_{3/2}(x)}{\,}\right]^2}.
\end{equation}

From the confinement phase, we immediately jump to the opposite limit of very high temperatures, $T\gg T_c$,
where the excitations of the second type are supposed to be mostly important.
There, the shear viscosity has the form~\cite{amy} 
\begin{equation}
\label{kn}
\eta=\frac{T^3}{g^4}{\,}\frac{27.126}{\ln\frac{2.765}{g}}.
\end{equation}
At such temperatures~\cite{nik}, $G_T\sim (g^2T)^4$, so that Eq.~(\ref{eta}) yields
$N'\sim\bigl(\frac{m}{g^4T}\bigr)^5$. [We use the notation $N'$ to make a distinction from Eq.~(\ref{N}).]
In order for this coefficient $N'$ to be constant, one should have 
$m\sim g^4T$. Thus, the known high-temperature expression for the shear viscosity can indeed be reproduced within a model of nonperturbative stochastic fields
with the correlation length $\sim (g^4T)^{-1}$.

To distinguish the two scales, $\sim g^2T$ and $\sim g^4T$, we use  
from now on the notations $M$ and $m$, respectively. Thus, 
\begin{equation}
\label{mM}
M=m_0f_T~~~ {\rm and}~~~ m=g^2M.
\end{equation}
Here, the continuous function $f_T$ can be chosen in the following form: 
\begin{equation}
\label{fT}
f_T= \left\{\begin{array}{rcl}\Bigl[\coth\Bigl(\frac{m_0}{4T}\Bigr)\Bigr]^{1/4}~~ {\rm at}~~ T_c<T<T_{*},\\
\Bigl[\coth\Bigl(\frac{m_0}{4T_{*}}\Bigr)\Bigr]^{1/4}\cdot
\frac{g^2{\,}T}{g^2_{*}{\,}T_{*}}~~ {\rm at}~~ T>T_{*}.\end{array}\right.
\end{equation}
The coth-factors in Eq.~(\ref{fT}) stem from the above-mentioned parametrization  
$D(x)={\rm e}^{-m_0|x|/2}$, which yields~\cite{nik}
$\frac{G_T}{G}=f_T^4=\coth\Bigl(\frac{m_0}{4T}\Bigr)$ at $T_c<T<T_{*}$. Furthermore, 
$T_{*}$ in Eq.~(\ref{fT}) is the temperature of dimensional reduction, and $g_{*}\equiv g(T_{*})$. Lattice simulations~\cite{f1} and analytic calculations~\cite{nik, yus1} 
suggest the value of $T_{*}$ in the range from $T_c$ to $2T_c$.
Below, we will find $T_{*}$ from this range by using the best known lattice value for the shear-viscosity to the entropy-density ratio.

We assume now the function $D(x)$ at $T>T_c$ in the form of a sum
\begin{equation}
\label{fu}
D(x)=D_M(x)+D_m(x).
\end{equation} 
Here, $D_M$ is given by Eq.~(\ref{d}) with $m_0$ replaced by $M$,
while $D_m$ is given by a similar formula: 
$D_m(x)=24{\,}\left[N'{\,}\frac{K_{3/2}(m|x|)}{(m|x|)^{3/2}}\right]^{1/2}$. Since $M\gg m$ at $T\gg T_c$,
one has at such temperatures $D\simeq D_m$ with an exponentially high accuracy. Therefore, the shear viscosity~(\ref{eta}) goes at $T\gg T_c$ as 
$$\eta\simeq\sqrt{8\pi^5}{\,}N'{\,}\frac{G_T^2}{m^5}\simeq
\sqrt{8\pi^5}{\,}N'{\,}\frac{G^2}{m_0^5}\cdot\frac{T^3}{g^4}\left\{\frac{
\Bigl[\coth\Bigl(\frac{m_0}{4T_{*}}\Bigr)\Bigr]^{1/4}}{g_{*}^2T_{*}}\right\}^3.$$
Comparing this expression with the known one, Eq.~(\ref{kn}), we get the coefficient $N'$:
$$N'=\frac{1}{\sqrt{8\pi^5}}{\,}\frac{27.126}{\ln\frac{2.765}{g}}{\,}
\left\{\frac{g_{*}^2T_{*}}{\Bigl[\coth\Bigl(\frac{m_0}{4T_{*}}\Bigr)\Bigr]^{1/4}}\right\}^3
\cdot\frac{m_0^5}{G^2}.$$ 
As it was anticipated above, this result is $T$-independent with the double logarithmic accuracy (since 
$g$ depends on $T$ only logarithmically).

We can now proceed towards the main result of the present paper --- the full shear viscosity
produced by the two types of excitations of the nonperturbative stochastic fields, which are described by the correlation function~(\ref{fu}),
$$D(x)=24\left\{\left[N{\,}\frac{K_{3/2}(M|x|)}{(M|x|)^{3/2}}\right]^{1/2}+
\left[N'{\,}\frac{K_{3/2}(m|x|)}{(m|x|)^{3/2}}\right]^{1/2}\right\}.$$
As it follows from the above analysis, the full viscosity is given by the formula
\begin{equation}
\label{fu1}
\eta=\sqrt{8\pi^5}{\,}G_T^2\left(\frac{N}{M^5}+\frac{N'}{m^5}\right)+\Delta\eta,
\end{equation}
where the contribution $\Delta\eta$ is produced by the interaction between these two types of excitations.
This contribution corresponds to the cross term in the square of the function $D(x)$:
\begin{equation}
\label{ct}
{\rm cross~ term}=24^2\cdot 2\left[NN'{\,} \frac{K_{3/2}(M|x|)}{(M|x|)^{3/2}}
{\,}\frac{K_{3/2}(m|x|)}{(m|x|)^{3/2}}\right]^{1/2}.
\end{equation}
This cross term can be approximated by a function of the 
type of Eq.~(\ref{d}) as
\begin{equation}
\label{ct1}
{\rm cross~ term}\simeq 24^2\cdot {\cal N}{\,}\frac{K_{3/2}(\mu|x|)}{(\mu|x|)^{3/2}},
\end{equation}
where $\mu=\frac{M+m}{2}$. Then, by virtue of Eq.~(\ref{eta}), $\Delta\eta$ can be expressed through the mass parameter $\mu$ and the yet unknown coefficient ${\cal N}$ as 
\begin{equation}
\label{de}
\Delta\eta=\sqrt{8\pi^5}{\,}{\cal N}{\,}\frac{G_T^2}{\mu^5}.
\end{equation} 
In terms of the spectral density, this means that the interaction between the two types of excitations
is also modeled by the Lorentzian Ansatz~(\ref{ans}), whose width $\mu$ is given by 
the mean value of $M$ and $m$.

Using now the explicit form of the function 
$\frac{K_{3/2}(z)}{z^{3/2}}$, which can be found in Eq.~(\ref{uu}), we see that the approximation~(\ref{ct1})
to Eq.~(\ref{ct}) can be written as follows:
$${\cal N}^2\simeq\frac{4NN'\mu^4}{M^2m^2}{\,}
\frac{\left(\frac{1}{M|x|}+1\right)\left(\frac{1}{m|x|}+1\right)}{\left(\frac{1}{\mu|x|}+1\right)^2}.$$
To make the right-hand side of this expression really constant, we have to 
disregard ``+1'' in all the three brackets. That is, we restrict ourselves to the leading 
pre-exponential terms 
in the exponentially falling off functions $\frac{K_{3/2}(M|x|)}{(M|x|)^{3/2}}$, 
$\frac{K_{3/2}(m|x|)}{(m|x|)^{3/2}}$, and $\frac{K_{3/2}(\mu|x|)}{(\mu|x|)^{3/2}}$. This yields the 
following result:
$${\cal N}\simeq 2\sqrt{NN'}{\,}\frac{\mu^3}{(Mm)^{3/2}}.$$
Owing to Eq.~(\ref{de}), the sought full shear viscosity~(\ref{fu1}) finally reads
\begin{equation}
\label{et5}
\eta\simeq
\sqrt{8\pi^5}{\,}G_T^2\left[\frac{N}{M^5}+\frac{N'}{m^5}+\frac{2\sqrt{NN'}}{\mu^2(Mm)^{3/2}}\right].
\end{equation}
At $T\gg T_c$, the three terms on the right-hand side of Eq.~(\ref{et5}), once divided by $T^3$, scale (with the double logarithmic accuracy) as ${\cal O}(g^6)$, ${\cal O}(1/g^4)$, and ${\cal O}(g^3)$, respectively. We see that, at such temperatures, $\Delta\eta$ is parametrically larger than the contribution of excitations with the correlation length $1/M$ by a factor of ${\cal O}(1/g^3)$, while being parametrically smaller than the contribution of excitations with the correlation length $1/m$ by a factor of ${\cal O}(g^7)$.

\begin{figure}
\psfrag{P}{\Large{$s/T^3$}}
\epsfig{file=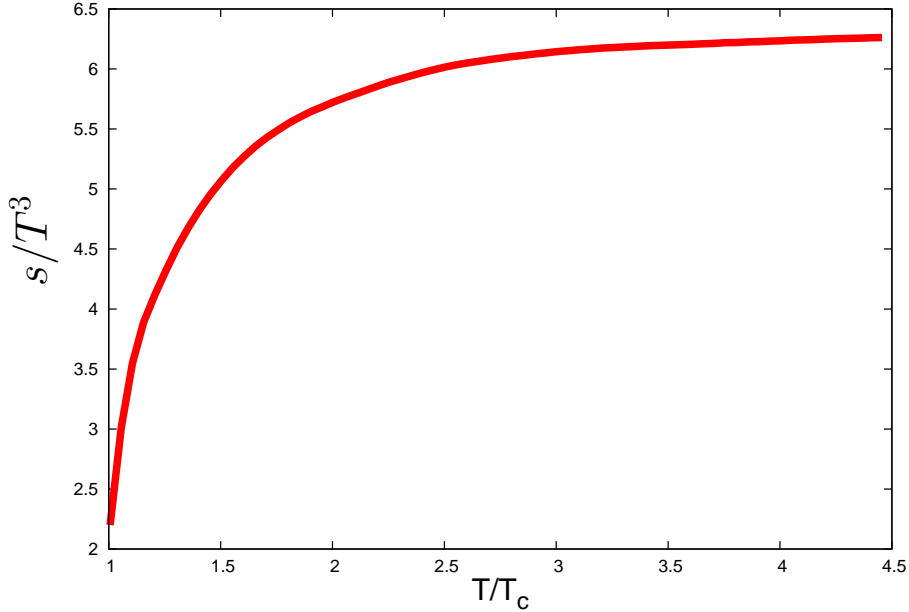, width=120mm}
\caption{\label{222}The $(s/T^3)$-ratio corresponding to the lattice values 
for the pressure from Ref.~\cite{f1} (courtesy of F.~Karsch), as a function of $T/T_c$.}
\end{figure}

In order to evaluate Eq.~(\ref{et5}) numerically, we adopt the lattice-simulated 
SU(3) Yang--Mills critical temperature
$T_c=270{\,}{\rm MeV}$ and the two-loop running coupling~\cite{f1} 
$$g^{-2}=2b_0\ln\frac{T}{\Lambda}+\frac{b_1}{b_0}\ln\left(2\ln\frac{T}{\Lambda}\right), {\rm where}~
b_0=\frac{11N_c}{48\pi^2},{\,} b_1=\frac{34}{3}\left(\frac{N_c}{16\pi^2}\right)^2,{\,} \Lambda=0.104T_c,{\,} N_c=3.$$
For the correlation length $\frac{m_0}{2}$ of the SU(3) Yang--Mills 
vacuum in the confinement phase, entering Eq.~(\ref{mM}),
we use the lattice value from Ref.~\cite{lat}, which reads
$\frac{m_0}{2}=894{\,}{\rm MeV}$. Furthermore, also in the confinement phase, 
we should choose such a 
lattice value of the gluon condensate $G$ that corresponds to an Ansatz for the two-point correlation function~(\ref{p}) without the $\frac{1}{|x|^4}$-term. According to Ref.~\cite{en}, this value is 
$G=2.21{\,}{\rm GeV}^4$. 
Next, the entropy density, $s=s(T)$,
in which units $\eta$ is measured, can be obtained by the formula
$s=dp_{\rm lat}/dT$, where we use for the pressure $p_{\rm lat}$ 
the corresponding lattice values from Ref.~\cite{f1}. 
The value $\frac{s(T_c)}{T_c^3}\simeq 2.2$ is fixed by imposing cancellation of the latent-heat contribution 
to the pressure. Namely, this contribution is expressed by the 
discontinuity of $\varepsilon/T^4$ at $T=T_c$, which, according to Fig.~7 of Ref.~\cite{f1}, is $\simeq2.2$.
In Fig.~\ref{222}, we plot the resulting $s/T^3$ as a function of $T/T_c$.

Implying the subtraction of the vacuum contribution, and noticing that $\coth\Bigl(\frac{m_0}{4T_c}\Bigr)=1.07\simeq1$,
we use in Eq.~(\ref{et5}) a parametrization of $G_T$ in the form $G_T=G\cdot (f_T^4-1)$, where 
$f_T$ is given by Eq.~(\ref{fT}). We find the value of $T_{*}$ upon the comparison with the best known lattice value of the $\frac{\eta}{s}$-ratio~\cite{me}, which reads $\frac{\eta}{s}\bigr|_{T=1.65{\,}T_c}=0.134$. Then, the value 
of $T_{*}$ comes out to be just $T_{*}\simeq1.65{\,}T_c$, and the resulting $\frac{\eta}{s}$ turns out to be a monotonically 
increasing function with the minimum value at $T=T_c$: $\frac{\eta}{s}\bigr|_{T=T_c}\simeq 0.081$. We notice that
this value is only a few per cent larger than the lower bound of $\frac{1}{4\pi}$ predicted by the AdS/CFT-correspondence~\cite{pss}. The resulting 
$\frac{\eta}{s}$-ratio is plotted in Fig.~\ref{2t}. Also, using this 
value of $T_{*}$,
we plot in Fig.~\ref{3t} the ratios of the 1st and the 3rd terms 
on the right-hand side of Eq.~(\ref{et5}) to the leading, 2nd, term. As mentioned above, the first of these two ratios goes as ${\cal O}(g^{10})$, becoming at $T=4.5{\,}T_c$ as small as only 0.01, whereas the second ratio
goes as ${\cal O}(g^7)$.

Furthermore, 
we evaluate the extent to which the correcting 
factor~(\ref{f3}) at $|k|\sim 1$ may differ from 1 for temperatures $T\in(T_c,T_{*})$.
It turns out that, regardless
of the argument, $m$, $M$, or $\mu$, entering this correcting factor, the maximum value it can reach is
$<1.42$. Consequently, at $T\in(T_c,T_{*})$, the resulting $\frac{\eta}{s}$ 
may be up to 42\% larger. That would lead to a discontinuity at $T=T_{*}$
of the curve depicted in Fig.~\ref{2t}, since
in the dimensionally-reduced phase of $T>T_{*}$ the correcting 
factor~(\ref{f3}) is always equal to 1. This uncertainty, emerging due to possible deviations of the correcting factor from 1 at $T\in(T_c,T_{*})$, is related to a nonuniversal character of the Lorentzian Ansatz~(\ref{ans}).

\section{Concluding remarks} 

\begin{figure}
\psfrag{y}{\Large{$\eta/s$}}
\epsfig{file=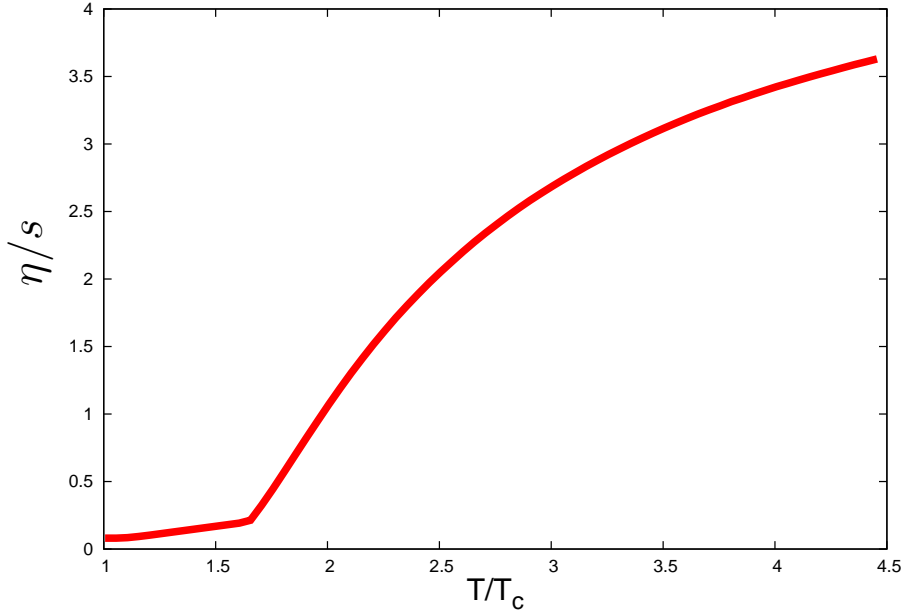, width=120mm}
\caption{\label{2t}The evaluated $(\eta/s)$-ratio as a function of $T/T_c$.}
\end{figure}

\begin{figure}
\epsfig{file=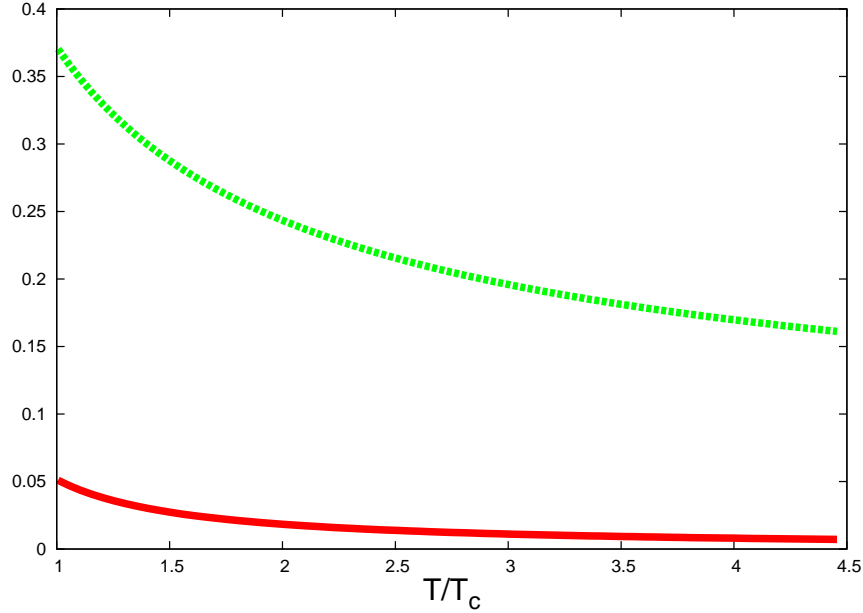, width=120mm}
\caption{\label{3t}The ratios of the 1st and the 3rd terms 
on the right-hand side of Eq.~(\ref{et5}) to the 2nd term are represented by the lower and the upper curves,
respectively.}
\end{figure}

In this Letter, we have considered shear viscosity within a model of the gluon plasma based entirely on the nonperturbative stochastic 
fields which, however, may develop excitations with two different correlation lengths, $\sim (g^2T)^{-1}$ and 
$\sim (g^4T)^{-1}$. Excitations of the first type possess all the features of the genuine nonperturbative stochastic chromo-magnetic fields, which survive the deconfinement phase transition in Yang--Mills theory~\cite{lat}. Rather, excitations of the second type are 
introduced for the sole purpose to reproduce,
within one and the same model, the known high-temperature perturbative contribution to the shear viscosity.
Bringing both types of excitations together, we have studied the contribution of nonperturbative stochastic 
Yang--Mills fields to the shear viscosity, relative to the perturbative contribution, as well as the result of 
the perturbative-nonperturbative interference (cf. Fig.~\ref{3t}). Remarkably, the interference contribution parametrically exceeds the purely nonperturbative one by a factor of ${\cal O}(1/g^3)$. Furthermore, using the best known lattice value for the $\frac{\eta}{s}$-ratio from Ref.~\cite{me}, we have evaluated the temperature dependence of this ratio, which is  depicted in Fig.~\ref{2t}. The minimum value of the resulting $\frac{\eta}{s}$-ratio, reached at
$T=T_c$, turns out to be only a few per cent larger than the 
lower bound of $\frac{1}{4\pi}$ predicted for this quantity by the AdS/CFT-correspondence.

\acknowledgments

\noindent
The author is grateful to E.~Meggiolaro, H.-J.~Pirner, and J.E.F.T. Ribeiro for the useful discussions, and to 
F.~Karsch for providing the details of the lattice data from Ref.~\cite{f1}. 
This work was supported by the Portuguese Foundation for Science and Technology
(FCT, program Ci\^encia-2008) and by 
the Center for Physics of Fundamental Interactions (CFIF) at Instituto Superior
T\'ecnico (IST), Lisbon.

\end{document}